\documentclass[twoside,twocolumn,9pt]{article}
\usepackage{extsizes}
\usepackage[super,sort&compress,comma]{natbib} 
\usepackage[version=3]{mhchem}
\usepackage[left=1.5cm, right=1.5cm, top=1.785cm, bottom=2.0cm]{geometry}
\usepackage{balance}
\usepackage{mathptmx}
\usepackage{sectsty}
\usepackage{graphicx} 
\usepackage{lastpage}
\usepackage[format=plain,justification=justified,singlelinecheck=false,font={stretch=1.125,small,sf},labelfont=bf,labelsep=space]{caption}
\usepackage{float}
\usepackage{fancyhdr}
\usepackage{fnpos}
\usepackage[english]{babel}
\addto{\captionsenglish}{%
  
}
\usepackage{array}
\usepackage{droidsans}
\usepackage{charter}
\usepackage[T1]{fontenc}
\usepackage[usenames,dvipsnames]{xcolor}
\usepackage{setspace}
\usepackage[compact]{titlesec}
\usepackage{hyperref}
\usepackage{amsmath}
\usepackage{amsfonts}
\usepackage{amssymb}
\usepackage{ulem}

\usepackage{epstopdf}%This line makes .eps figures into .pdf - please comment out if not required.

\definecolor{cream}{RGB}{222,217,201}

\begin{document}

\pagestyle{fancy}
\thispagestyle{plain}
\fancypagestyle{plain}{
%%%HEADER%%%
\renewcommand{\headrulewidth}{0pt}
}
%%%END OF HEADER%%%

%%%PAGE SETUP - Please do not change any commands within this section%%%
\makeFNbottom
\makeatletter
\renewcommand\LARGE{\@setfontsize\LARGE{15pt}{17}}
\renewcommand\Large{\@setfontsize\Large{12pt}{14}}
\renewcommand\large{\@setfontsize\large{10pt}{12}}
\renewcommand\footnotesize{\@setfontsize\footnotesize{7pt}{10}}
\makeatother

\renewcommand{\thefootnote}{\fnsymbol{footnote}}
\renewcommand\footnoterule{\vspace*{1pt}% 
\color{cream}\hrule width 3.5in height 0.4pt \color{black}\vspace*{5pt}} 
\setcounter{secnumdepth}{5}

\makeatletter 
\renewcommand\@biblabel[1]{#1}            
\renewcommand\@makefntext[1]% 
{\noindent\makebox[0pt][r]{\@thefnmark\,}#1}
\makeatother 
\renewcommand{\figurename}{\small{Fig.}~}
\sectionfont{\sffamily\Large}
\subsectionfont{\normalsize}
\subsubsectionfont{\bf}
\setstretch{1.125} %In particular, please do not alter this line.
\setlength{\skip\footins}{0.8cm}
\setlength{\footnotesep}{0.25cm}
\setlength{\jot}{10pt}
\titlespacing*{\section}{0pt}{4pt}{4pt}
\titlespacing*{\subsection}{0pt}{15pt}{1pt}
%%%END OF PAGE SETUP%%%

%%%FOOTER%%%
\fancyfoot{}
%\fancyfoot[LO,RE]{\vspace{-7.1pt}\includegraphics[height=9pt]{head_foot/LF}}
%\fancyfoot[CO]{\vspace{-7.1pt}\hspace{13.2cm}\includegraphics{head_foot/RF}}
%\fancyfoot[CE]{\vspace{-7.2pt}\hspace{-14.2cm}\includegraphics{head_foot/RF}}
\fancyfoot[RO]{\footnotesize{\sffamily{1--\pageref{LastPage} ~\textbar  \hspace{2pt}\thepage}}}
\fancyfoot[LE]{\footnotesize{\sffamily{\thepage~\textbar\hspace{3.45cm} 1--\pageref{LastPage}}}}
\fancyhead{}
\renewcommand{\headrulewidth}{0pt} 
\renewcommand{\footrulewidth}{0pt}
\setlength{\arrayrulewidth}{1pt}
\setlength{\columnsep}{6.5mm}
\setlength\bibsep{1pt}
%%%END OF FOOTER%%%

%%%FIGURE SETUP - please do not change any commands within this section%%%
\makeatletter 
\newlength{\figrulesep} 
\setlength{\figrulesep}{0.5\textfloatsep} 

\newcommand{\topfigrule}{\vspace*{-1pt}% 
\noindent{\color{cream}\rule[-\figrulesep]{\columnwidth}{1.5pt}} }

\newcommand{\botfigrule}{\vspace*{-2pt}% 
\noindent{\color{cream}\rule[\figrulesep]{\columnwidth}{1.5pt}} }

\newcommand{\dblfigrule}{\vspace*{-1pt}% 
\noindent{\color{cream}\rule[-\figrulesep]{\textwidth}{1.5pt}} }

\makeatother
%%%END OF FIGURE SETUP%%%

%%%TITLE, AUTHORS AND ABSTRACT%%%
\twocolumn[
  \begin{@twocolumnfalse}
%{\includegraphics[height=30pt]{head_foot/journal_name}\hfill\raisebox{0pt}[0pt][0pt]{\includegraphics[height=55pt]{head_foot/RSC_LOGO_CMYK}}\\[1ex]
%\includegraphics[width=18.5cm]{head_foot/header_bar}}\par
\vspace{1em}
\sffamily
\begin{tabular}{m{4.5cm} p{13.5cm} }

 & \noindent\LARGE{Modeling electrodeposition of thin films with compact, dendritic, and random morphologies} \\%Article title goes here instead of the text "This is the title"
\vspace{0.3cm} & \vspace{0.3cm} \\

 & \noindent\large{F\'abio D. A. Aar\~ao Reis,\textit{$^{a}$} Dung di Caprio,\textit{$^{b}$} Sidiney G. Alves\textit{$^{c}$} and Ismael S. S. Carrasco\textit{$^{d}$}} \\%Author names go here instead of "Full name", etc.

& \noindent\normalsize{
We summarize previously published computer simulation studies of a metal electrodeposition model that considers diffusive cation flux, a fast reaction in contact with the electrode or with the growing deposit, and adsorbed atom diffusion in finite time windows.
%R1Q1
Simulations with a variety of parameters show compact, fractal, and dendritic deposits, as well as dynamic transitions between different morphologies during the growth of individual samples.
The dendritic shapes are characterized by hierarchical organizations in which branches of a given order reproduce the shape and crystalline orientation of the branches of the previous order, in contrast to the random branching of the fractal deposits formed by diffusion-limited aggregation (DLA) and related models.
The simulations show dendrites with maple leaf shape in simple cubic lattices, pine tree shape in face centered cubic lattices, and flower-like shape in hexagonal close packed lattices, with surfaces predominantly oriented in the lowest-energy configurations of each lattice.
The characteristic widths of the dendrites are roughly proportional to the diffusion lengths of the adsorbed atoms which are weakly bound to growing islands on the lowest-energy surfaces.
These morphologies reflect the competition between the relaxation that produces low-energy configurations at short length scales and the diffusive cation flux that privileges reaction at protuberant surface points and leads to unstable growth at long length scales.
Simulations with nucleation restricted to a nanoscale seed show the initial formation of faceted particles followed by dendrite formation, in agreement with silver electrodeposition on substrates patterned with gold nanoparticles.
A quantitative application to faceted and dendritic silver obtained in the same experiment provides values of the model parameters which are consistent with the characteristic widths of these structures being of the same order of magnitude of the diffusion lengths of atoms on (111) and (001) surfaces.
Using an extended model for iron electrodeposition on Si(100), the simulations qualitatively reproduce the initial formation of multilayer islands, small oscillations of the roughness of a polycrystalline compact film after island coalescence, and instability development.
%R1Q4
Possible directions for applying this modeling to the electrodeposition of other materials are suggested.} \\

\end{tabular}

 \end{@twocolumnfalse} \vspace{0.6cm}

  ]

\renewcommand*\rmdefault{bch}\normalfont\upshape
\rmfamily
\section*{}
\vspace{-1cm}

\footnotetext{\textit{$^{a}$~Instituto de F\'{i}sica, Universidade Federal Fluminense,
		Avenida Litor\^{a}nea s/n, 24210-340 Niter\'{o}i, RJ, Brazil; E-mail: fdaar@protonmail.com}}
\footnotetext{\textit{$^{b}$~PSL Research University, Chimie ParisTech - CNRS, Institut de Recherche de Chimie Paris, 75005, Paris, France; E-mail: dung.di-caprio@chimieparistech.psl.fr}}
\footnotetext{\textit{$^{c}$~Departamento de Estat\'istica, F\'isica e Matem\'atica, Universidade Federal de S\~ao Jo\~ao Del-Rei, 36420-000, Ouro Branco, MG, Brazil; E-mail: sidiney@ufsj.edu.br}}
\footnotetext{\textit{$^{d}$~Departamento de F\'isica, Universidade Federal de Vi\c cosa, 36570-000, Vi\c cosa, MG, Brazil; E-mail: theismiu@gmail.com}}

\section{Introduction}

Electrodeposition is a widely used technique for the production of thin films and other nanostructures due to its relatively low cost and its ability to produce materials with different shapes by fine-tuning the growth conditions \citep{gamburg,wangbook}.
This justified the development of several electrodeposition models, which may represent the dynamics of the chemical species in the electrolyte, their interaction with the surface where the relevant reactions occur, and the dynamics of the deposited material.
For metal electrodeposition, one of the simplest models is diffusion-limited aggregation (DLA) \citep{witten}, which forms highly porous deposits with randomly distributed branches \citep{witten,meakin1983,meakin1984}.
DLA models consider the sequential release of cations far from the growing deposit, their diffusive motion in the electrolyte, reduction of each cation at the first contact with the electrode or with the deposit, and permanent aggregation of the resulting metal atom at that position.
The diffusive cation motion is reasonable if a highly depleted region is formed near the deposit and if the electrolyte is supported, so that the electrostatic field and the consequent ion migration are suppressed.
The first contact reaction is expected when the reaction rate is much larger than the rate of diffusion at atomic length scale (i.e. a high Damkohler number at this scale \citep{fogler}).
Finally, the permanent aggregation of the reduced atom is an assumption that adsorbate relaxation is negligible.

The choice of supported electrolytes is common in electrodeposition practice \citep{gamburg,wangbook}, although it is important to recall that ramified (DLA-like) deposits may also be produced when ion migration is present, which can be explained by other modeling approaches \citep{chazalviel1990}.
The rapid cation reduction and the formation of a highly depleted region near the work electrode are somewhat related assumptions; conversely, a slow reaction might lead to a finite concentration at that region.
However, contrary to the DLA model assumption, there is evidence that diffusion of the adsorbed material plays a role in metal electrodeposition processes.
This is revealed by temperature effects on film morphology \citep{grunder,aryanfar2015}, investigations of the early growth stages \citep{berkesCPC2012,desai,nishikawa,ustarroz2013,xuJPCC2017}, transmission electron microscopy (TEM) studies of dendrite evolution \citep{fang2006,nishikawa,fang2008a,sunSciRep2013}, molecular dynamics simulations \citep{fang2008}, and scaling concepts applied to surface roughening \citep{PB,brandt2015}.

% R1Q6
These observations motivated several computational approaches for modeling electrodeposition.
Continuum descriptions such as phase-field formulations couple ionic transport and electrochemical kinetics by representing ion concentrations and interface motion as continuous fields \citep{cogswell2015,wang2020phasefield}.
The transport and chemical parameters are averaged over length scales much larger than the molecular sizes, so they do not describe details of the growing crystals and of molecular or ionic reactions.
At much smaller length and time scales, molecular dynamics and related atomistic methods explicitly represent elementary processes such as ion transport, reduction, and adsorbed atom (adatom) motion.
They advance in explaining nanoscale features of electrodeposits \citep{chaudhuri2025}, but the choice of interaction potentials involves approximations that may affect the results.
Moreover, the high computational cost of these methods generally restricts them to small system sizes and short growth times.
Between these limits, discrete Monte Carlo approaches, ranging from DLA extensions to more detailed kinetic Monte Carlo (KMC) models, use stochastic representations of particle deposition and surface diffusion on lattices, often sharing surface-relaxation mechanisms with models of deposition from vapor \citep{michely,etb}.
The simplified treatment of interfacial chemistry and atomic interactions allows a general assessment of how local rules shape morphology over larger length and time scales \citep{castro2000,liuEActa2013,tanyakarn,haoACSAMI2018,vishnugopi2020}.
KMC simulations reproduce some experimentally observed morphologies, such as the compact, columnar, porous, and fractal deposits, and represent transitions between these morphologies \citep{castro2000,haoACSAMI2018,vishnugopi2020}.
The fractal regime is obtained in conditions similar to DLAs: high energy barriers that effectively prevent surface diffusion and high overpotentials that lead to rapid deposition \citep{haoACSAMI2018}.

However, none of the above models show the formation of stable branched structures with hierarchical organizations, which are termed dendrites in the present work.
In a dendrite, branches of a given order reproduce the shape and crystalline orientation of the branches of the previous order but grow in different directions.
They are observed in the electrodeposition of materials such as gold \citep{yeEActa2010,yeJPCC2010}, cobalt \citep{monzonCM2012}, silver \citep{talebEActa2013a,liuCEC2015,cheng2019,amin2021}, palladium \citep{maniam2013}, and copper \citep{talebEActa2013}, with a variety of shapes.
These dendritic structures have advantageous physical and chemical properties, such as enhanced hydrofobicity \citep{yeJPCC2010,talebEActa2013a}, improved sensing in surface-enhanced Raman scattering (SERS) devices \citep{yeEActa2010,cheng2019,amin2021}, and increased electrocatalytic activity \citep{maniam2013}.

An attempt to represent dendrite growth in electrodeposition modeling was partly inspired by the conclusions of TEM investigations of silver and lead dendrites \citep{fang2006,sunSciRep2013}.
These studies showed that the dendrite tips have amorphous or polycrystalline layers with thicknesses of a few nanometers and showed that these layers relax until they are incorporated into the core crystals.
The dendrite stability indicates that there is no coarsening of the core crystals, i.e., they do not relax to configurations with lower energies (e.g., compact micro or nanoparticles).
This motivated an extension of the DLA model in which each adatom formed by the cation reduction executes surface diffusion in contact with a quenched crystal configuration during a finite time window \citep{reiscapriotaleb2017,capriotalebreis2018}.
The single adatom diffusion is an approximation for the observed relaxation of amorphous or polycrystalline layers in contact with the stable core crystal, while the permanent aggregation of the adatom after this process represents its incorporation into the stable crystal.
Computer simulations of the model showed the formation of several dendritic structures, which depend on the underlying lattice geometry [e.g., dendrites in simple cubic (SC) lattices and face-centered cubic (FCC) lattices are different] and on the electrode orientation, and showed dynamical transitions from compact to dendritic morphologies as the deposits grow.

%R1Q1
In this highlight paper, we present details of this model (Sec. \ref{model}), review the previous findings on the qualitative and quantitative effects of the model parameters on the dendrite geometry and on the transitions to other geometries (Sec. \ref{results}), presents previous applications to silver and iron electrodeposition (Sec. \ref{applications}), and discuss some directions in which this modeling approach may be extended together with our concluding remarks (Sec. \ref{conclusion}).

%R1minor2
\begin{table}[t]
	\centering
	\caption{List of acronyms and symbols}
	\label{tab_acronyms}
	\small
	\setlength{\tabcolsep}{3pt}
	\begin{tabular}{@{}p{0.19\columnwidth}p{0.75\columnwidth}@{}}
		\hline
		Symbol & Meaning \\
		\hline
		AFM  & Atomic force microscopy \\
		BCC  & Body-centered cubic \\
		DLA  & Diffusion-limited aggregation \\
		FCC  & Face-centered cubic \\
		HCP  & Hexagonal close-packed \\
		HOPG & Highly oriented pyrolytic graphite \\
		KMC	 & kinetic Monte Carlo \\
		ML   & Monolayer \\
		NN   & Nearest neighbor (lattice dependent) \\
		NNN  & Next-nearest neighbor (lattice dependent) \\
		NP   & Nanoparticle \\
		SC   & Simple cubic \\
		SEM  & Scanning electron microscopy \\
		SERS & Surface-enhanced Raman scattering \\
		TEM  & Transmission electron microscopy \\[2pt]
		$A$ & Area of connected domain in dendrite cross section \\
		$E_G$, $E_P$ & Effective activation energies associated with $G$ and $P$ \\
		$G$ & Parameter defining the number of hop attempts \\
		$G_0$ & Constant prefactor of $G$ \\
		$G_s$, $G_f$ & Hop limits on the Si substrate and Fe film \\
		$I$ & Applied current \\
		$k$, $m$ & Exponents of relating characteristic width, $G$, and $P$ \\
		$k_{\mathrm{B}}$ & Boltzmann constant \\
		$l$ & AFM image or gliding-box size \\
		$n$ & Number of occupied NN sites \\
		$P$ & Parameter controlling the hop probability \\
		$P_{\mathrm{hop}}$ & Hop execution probability \\
		$S_{\mathrm{exp}}$ & Experimental microparticle facet size \\
		$S_t$ & Compact-particle size at the morphology transition \\
		$T$ & Absolute temperature \\
		$t$ & Dimensionless simulation time \\
		$W$ & Characteristic width of dendrites\\
		$W_{\mathrm{exp}}$ & Experimental nanotree characteristic size \\
		$w$ & Surface roughness \\
		$z_c$ & Compact layer thickness \\
		$\gamma$ & Exponent relating compact layer thickness and $G$ \\
		$\xi$ & Lateral correlation length \\
		\hline
	\end{tabular}
\end{table}

\section{Electrodeposition model with adsorbed atom diffusion}
\label{model}

\begin{figure*}%[!h]
	\centering
	\includegraphics[width=0.9\textwidth]{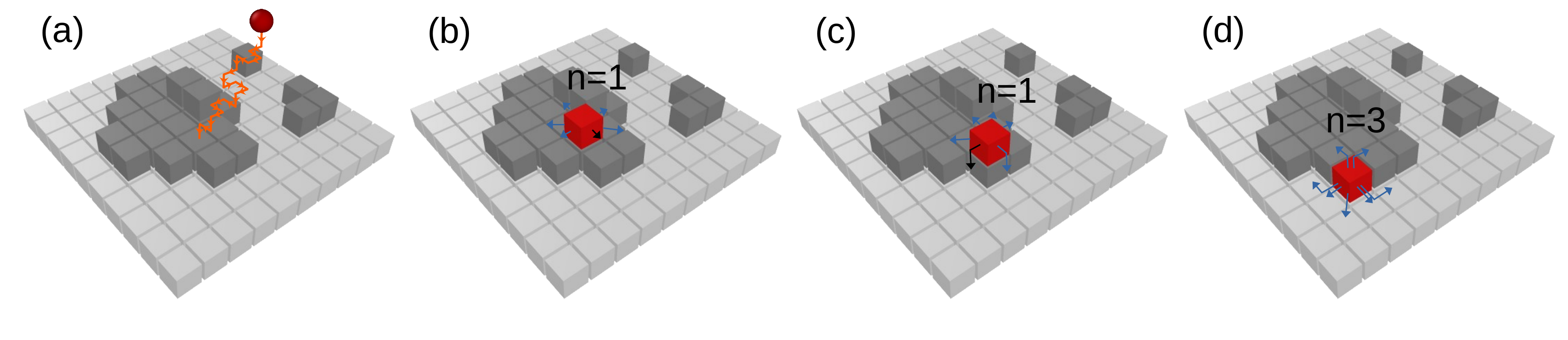}
	\caption{Deposition of an adatom on a given configuration of the previously formed deposit (dark gray) and of the electrode (light gray):
		(a) Random walk of a cation (dark red sphere) in the electrolyte.
		(b)-(d) First three hops of the mobile adatom formed by the cation reaction, with possible hop directions indicated by arrows [black arrows indicate the executed hops in (b) and (c)].}
	\label{figmodel}
\end{figure*}

%R1Q2
The steps of the model from the motion of a cation to the incorporation of an atom into the immobile crystal are illustrated in Fig. \ref{figmodel} for the case of a deposit with SC structure \citep{reiscapriotaleb2017,capriotalebreis2018}.
The (001) orientation is chosen for the planar electrode.
Each cation is released at a chosen distance (typically $30\text{--}50$ lattice units) above the topmost point of the deposit, which accounts for the formation of a region of the electrolyte with depleted concentration.
The cation then executes a random walk and reacts when it has a nearest neighbor (NN) site belonging to the electrode or containing an aggregated atom; the latter case is shown in Fig. \ref{figmodel}(a).

The product of the reaction is a mobile adatom which attempts to hop $G$ times to NN or next nearest neighbor (NNN) sites, where $G$ is the first model parameter.
In each attempt, the hop probability is
\begin{equation}
	P_{hop}=P^{n-1} ,
	\label{Phop}
\end{equation}
where $P<1$ is the second model parameter and $n$ is the number of occupied NN sites at the current position.
Moreover, the hop is executed only if the target site (randomly chosen among NNs and NNNs) has at least one NN site of the deposit or of the electrode.
The first three executed hops and the corresponding values of $n$ are shown in Figs. \ref{figmodel}(b)-(d).
The adatom is incorporated into the crystal at the final position after this sequence of hop attempts (all hop attempts count toward the limit of $G$ hops, including the rejected ones).
%R1minor5

%R1minor2
The model in FCC and hexagonal (HCP) lattices was defined as above, except that adatom hops were allowed only to NN sites of these lattices (unlike the SC lattice, step edges can be crossed by NN hops in FCC and HCP lattices).
In FCC lattices, simulations were also performed with non planar electrodes \citep{dokhan2022}, such as electrodes with a single faceted nanoparticle, with patterns of nanoparticles \citep{taleb2011}, and with large faceted particles.

%R1minor5
% R1Q2
The prescribed conditions for adatom hops simplify the evolution of amorphous or polycrystalline layers observed in TEM studies of silver and lead dendrites \citep{fang2006,sunSciRep2013}.
These layers reorganize in a finite time interval (not accessed in the TEM studies) before becoming permanently incorporated into the stable crystalline core.
The limitation to $G$ allowed hops for each adatom is consistent with the adatom incorporation into a stable crystal after that finite time interval, while the parameter $P<1$ helps the incorporation at positions with the lowest energies [highest $n$; Eq. (\ref{Phop})].
The main simplification is the assumption of single adatom diffusion to mimic the effects of the relaxation of many interacting adatoms in the layers surrounding the crystalline cores.
Thus, this electrodeposition model is designed as a minimal model to represent the competition between diffusive cation flux and a finite-time surface relaxation that produce stable compact or dendritic structures. 

The simulations of the model keep $G$ and $P$ constant as the deposit grows, which means that the hopping times and hop probabilities remain the same.
This is reasonable for a nearly constant deposition rate, which corresponds to nearly constant applied current $I$.
A current increase is expected to be accompanied by a higher crystallization rate, so we expect a shorter time for each adatom to move near the crystal surface.
Thus, $G$ is expected to decrease as $I$ increases.
The probability of an adatom to hop at a given surface configuration is expected to increase with the temperature $T$, so we expect that both $G$ and $P$ increase as $T$ increases.
For these reasons, keeping $G$ and $P$ constant when the deposit grows is suitable for modeling nearly galvanostatic deposition at constant temperature.

%R1Q3
In SC lattices, an improved version of the model considered that \citep{carrascoPCCP2022}: (i) the number of allowed hops $G$ is reset when the adatom moves away from a position with $n\geq2$ (i.e., more than one NN); (ii) the values of $G$ are extracted from Gaussian distributions.
The first change removes the memory of the surface diffusion on past configurations and allows a consistent interpretation of the parameter $P$ \citep{tosousareis2018}, while the second change introduces noise in the allowed surface diffusion time.
In FCC and HCP lattices, no such changes were implemented. 
The improved model produces dendrites with the same shapes as the original model and the dendrite sizes increased with both $G$ and $P$.
This led to the suggestion to write the parameters $G$ (average value) and $P$ as thermally activated quantities \citep{carrascoPCCP2022}:
\begin{equation}
	G \equiv \frac{G_0}{I^\alpha} \exp{\left( -\frac{E_G}{k_BT}\right)} \qquad , \qquad 
	P \equiv \exp{\left( -\frac{E_P}{k_BT}\right)} ,
	\label{defGP}
\end{equation}
where $G_0$ is a constant, $E_G>0$ and $E_P>0$ are effective activation energies, $\alpha>0$ is a dimensionless exponent, and $k_B$ is the Boltzmann constant.
However, the validity of these relations was not experimentally checked because it would depend on measuring data for a given material at several currents and temperatures.

%R2Q3
% R1Q2
The adatom diffusion mechanisms of this model differ from those of vapor deposition models \citep{etb,michely}.
The latter assume collective diffusion of all surface atoms with orientation dependent rates and assume energy barriers for adatom hops across step edges, which are expected to approximate the actual motion of individual adatoms.
However, the present model is not intended to represent details of the motion of the electrodeposited atoms, but to represent the consequences the relaxation of amorphous or polycrystalline layers before crystallization.

%R1Q2
Besides the above-mentioned approximations in the description of diffusive cation flux and adsorbate relaxation, other limitations of the model must also be stressed.
First, it omits coupled concentration and electric-potential fields considered in more detailed transport descriptions \citep{castro2000,haoACSAMI2018,vishnugopi2020}, with reaction chemistry reduced to simple first-contact rules.
Thus, the model is expected to be most informative when a supporting electrolyte suppresses migration and when reduction is fast compared to transport across nanoscale distances.
Second, the hopping rules assume isotropic interactions [Eq. (\ref{Phop})], which always privilege local configurations with the largest numbers of NNs, while the actual aggregation of adatoms or nanoclusters may have a more complex dependence on the crystalline orientation [as observed e.g. in vapor deposited films \citep{etb,michely}].
Accordingly, the model should be used to test plausible morphology-selection mechanisms and qualitative behavior, but  quantitative prediction of the evolution of specific materials may be more difficult; indeed, such predictions were already possible only in a small number of systems \citep{dokhan2022,benetti2023}.

Simulations of the model usually consider $P\geq0.001$, otherwise the typical hop probabilities are very small (particularly in FCC and HCP lattices) and the model becomes equivalent to DLA.
The values of $G$ range from $\sim10$ to $\sim{10}^6$ in most simulations in cubic cells of lateral width $\sim1500$; $G\sim{10}^8$ was already simulated in much smaller cells and much shorter times.
The cell widths were large enough to avoid finite size effects.
The time unit of the simulations usually corresponds to the aggregation of one monolayer of atoms.

\section{Dendrite growth and transitions to other morphologies}
\label{results}

\subsection{Simulations in SC lattices}

\begin{figure*}[!ht]
	\centering
	\includegraphics[width=0.85\textwidth]{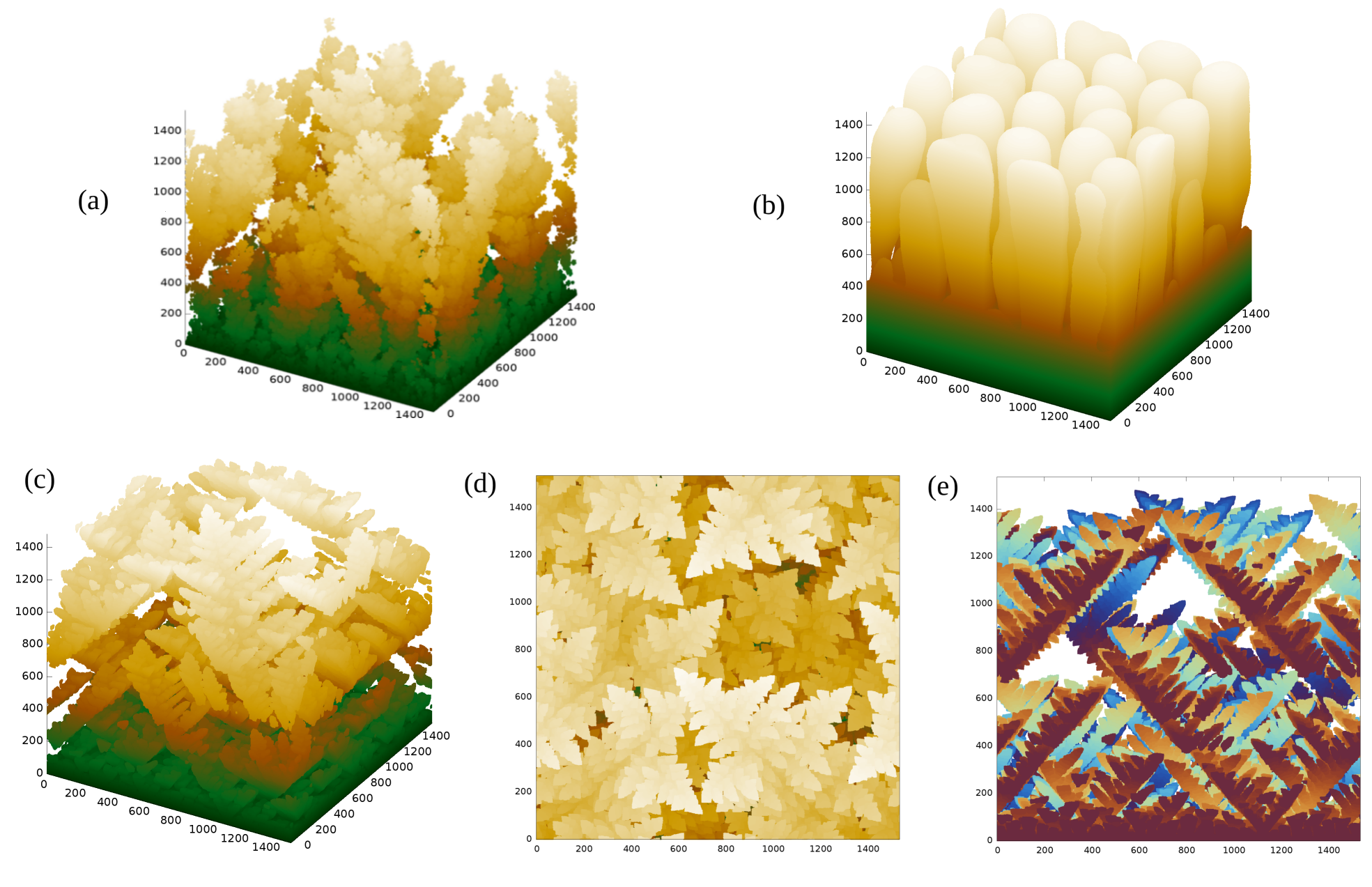}
	\caption{Deposits obtained in simulations on SC lattices with: (a) $G=20$ and $P=0.05$; (b) $G=5\times{10}^4$ and $P=0.25$; (c), (d), (e): $G=2\times{10}^3$ and $P=0.05$ with different viewpoints.
		Adapted with permission from Ref. \citenum{capriotalebreis2018}. Copyright 2018 American Chemical Society.}
	\label{examplesSC}
\end{figure*}

Small values of $G$ produce randomly branched deposits similar to those of DLA because the effect of the adsorbate diffusion is negligible, even if $P$ is not too small \citep{capriotalebreis2018}; see Fig. \ref{examplesSC}(a).
If $G$ is large (typically $\geq{10}^4$) and $P\gtrsim0.1$, a thick compact layer is initially formed; see the green region in Fig. \ref{examplesSC}(b).
However, surface fluctuations eventually lead to mound formation and subsequent growth of rounded columns \citep{reiscapriotaleb2017}, as shown in the yellow region of Fig. \ref{examplesSC}(b).
This shows that compact configurations are favored by the adatom motion to long distances, although an instability that leads to mound (or dendrite) formation eventually develops.
A scaling approach predicted that the instability develops when the velocity of propagation of lateral correlations becomes smaller than the upwards growth velocity \citep{reiscapriotaleb2017}.
This led to an estimate of the compact layer thickness 
\begin{equation}
	z_c\sim G^{\gamma} \quad ,\quad \gamma\approx 0.71 ,
	\label{zc}
\end{equation}
where the exponent $\gamma$ is related to the dynamical exponent of lateral correlations of interfaces whose growth is controlled by the Villain-Lai-Das Sarma equation \citep{villain,laidassarma}.
Simulation results were in excellent agreement with this prediction \citep{reiscapriotaleb2017}.

For intermediate values of $G$ and $P$, a narrow compact layer may be formed and the instability gives rise to dendrite growth, as shown in Figs. \ref{examplesSC}(c)-(e) from three different viewpoints.
Similar shapes are obtained for $P=0.05$ with other values of $G$ in the range ${10}^2\geq G\geq5\times{10}^4$; as $P$ increases, the range of $G$ that leads to dendrite formation broadens \citep{capriotalebreis2018}.

The dendrites have a maple leaf shape when viewed from the top [Fig. \ref{examplesSC}(d)].
Their tips are formed by three concurrent low-energy terraces ($xy$, $yz$, and $xz$ planes) and propagate in $\left[\pm1,\pm1,1\right]$ directions ($-z$ direction is excluded because cations are released at $z$ larger than the film heights).
The dendrite surface is mostly formed by the lowest-energy planes of the SC lattice [(001) and equivalent surfaces], which is expected because the adatom diffusion forms low-energy configurations at short length scales.
However, the time scale of the adatom diffusion is limited and, consequently, the surface diffusion length is limited.
Since the diffusive cation flux favors the aggregation at the most protuberant surface points (a tip effect), the instability develops (and dendrites are formed) at length scales that exceed that surface diffusion length.

\subsection{Simulations in FCC lattices}

Simulations in FCC lattices with small $G$ or small $P$ also lead to the formation of DLA-like morphologies, while compact columnar structures are also found with very large $G$ and $P\gtrsim0.3$, similarly to what was observed in SC lattices.
Intermediate values of $G$ and $P$ lead to dendrite formation.

\begin{figure}[!ht]
	\centering
	\includegraphics[width=0.33\textwidth]{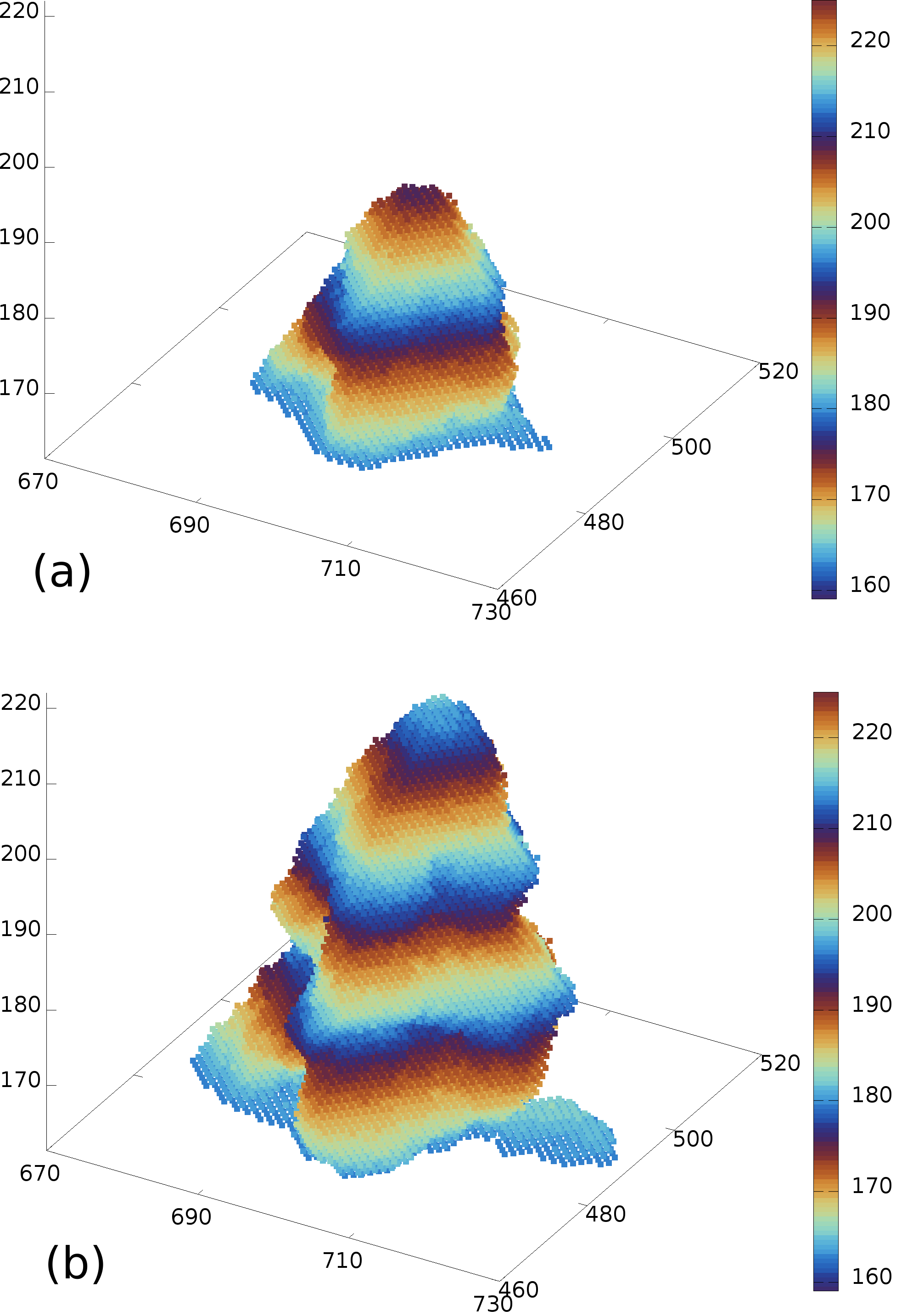}
	\caption{Snapshots of the initial growth with $G={10}^4$, $P=0.1$, and a small faceted nanoparticle in the origin, at the dimensionless simulation times (a) $t = 800$ and (b) $t = 1790$. Reproduced with permission from Ref. \citenum{dokhan2022}. Copyright 2022 American Chemical Society.}
	\label{initialFCC}
\end{figure}

%R1Q4 
Short-time simulations shown in Fig. \ref{initialFCC}(a) were obtained with $G={10}^4$, $P=0.1$, and a small faceted nanoparticle placed in the $z=0$ plane.
Correspondingly, the cation flux from top to bottom is predominant in the [0,0,$-1$] direction.
Despite these choices, the simulations show the formation of a compact particle with a pyramidal shape and dominant (111) surface, which is the lowest-energy surface of the FCC crystal.
As the growth proceeds, bifurcations appear in the middle of the pyramid [Fig. \ref{initialFCC}(b)], which will give rise to secondary branches that propagate in [$\pm1$,0,0] and [0,$\pm1$,0] directions, but whose facets predominantly have (111) or equivalent orientations.
These simulations show that finite-time adatom diffusion is consistent with the formation of low-energy surfaces at short length scales.
Importantly, the dominant (111) orientation is actually observed in faceted microparticles and dendrites of electrodeposited silver \citep{talebEActa2013a,jiang2012,radmilovic2016}.

\begin{figure*}[!ht]
	\centering
	\includegraphics[width=0.95\textwidth]{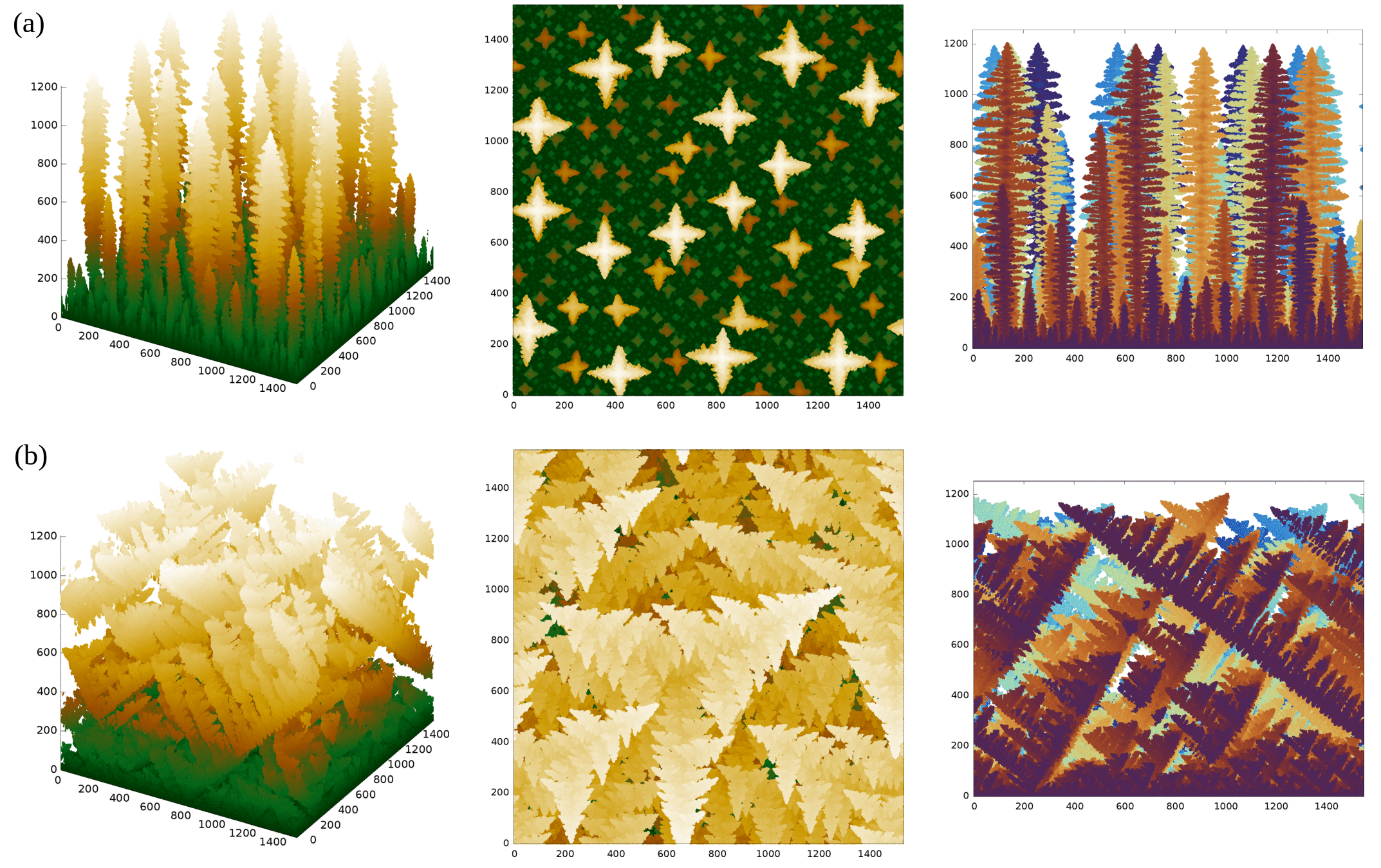}
	\caption{Deposits obtained in simulations on FCC lattices with $G=5\times{10}^4$ and $P=0.1$ with different viewpoints:
		(a) (001) substrate (cation flux in [0,0,$-1$] direction). Adapted with permission from Ref. \citenum{dokhan2022}. Copyright 2022 American Chemical Society.
		(b) (111) substrate (cation flux in [$-1$,$-1$,$-1$] direction). Adapted with permission from Ref. \citenum{capriotalebreis2025} (CC-BY 4.0 licence).
		}
	\label{examplesFCC}
\end{figure*}

Simulations with the same parameters but allowing nucleation at all points of the (001) electrode lead to growth of parallel dendrites with the shape of pine trees, as shown in Fig.~\ref{examplesFCC}(a) from three different viewpoints.
The dendrite trunks (primary branches) propagate in [001] direction and, at certain heights, there appear four secondary branches in [$\pm1$,0,0] and [0,$\pm1$,0] directions.
Small tertiary branches propagate in all these directions; the [0,0,$-1$] direction is excluded by the direction of the cation flux.

Fig. \ref{examplesFCC}(b) shows dendrite growth with the same model parameters but with an electrode with (111) orientation \citep{capriotalebreis2025}.
The electrode crystallography does not affect the dendrite shape, except possibly at very short times, but this orientation is coupled with the cation flux in the [$-1$,$-1$,$-1$] direction in the simulation cell.
The dendrites now grow with a leaf-like shape: the trunks propagate in [001], [010], and [100] directions; at some heights of the trunks, there appear only two secondary branches; a limitation in the number of tertiary branches is also observed.
This does not imply a change in the dendrite crystallography because (111) and equivalent planes are still the dominant surface directions.
The apparently different shape is a consequence of the direction of the cation flux, which suppresses the branches of the pine trees that would propagate in directions [$-1$,0,0], [0,$-1$,0], and [0,0,$-1$].
The lateral view of Fig. \ref{examplesFCC}(b) confirms that incipient secondary branches are formed in the opposite directions of the large secondary branches.

When the direction of the cation flux changes to [$-1$,$-1$,0], the dendrites also propagate in directions forming an acute angle with the electrode normal \citep{capriotalebreis2025}.
The appearance of pine trees is partially restored, but with incipient secondary branches in directions [$-1$,0,0] and [0,$-1$,0] \citep{capriotalebreis2025}.

The conclusion that can be derived from these results is that apparently different dendrite shapes may appear due to differences in the directions of the cation flux, but the crystallographic orientations do not change because they are correlated with the energetics of the adatom diffusion process.
More precisely, the present dendrite shapes are related to the assumption of isotropic interactions with the NNs [Eqs. (\ref{Phop}) and (\ref{defGP})], which favor formation of the lowest-energy surfaces (111) and branch propagation in directions which are only slightly more energetic [(001) and equivalent ones].
If different interactions are assumed (i.e., different energetics), then other dendrite shapes may be observed.
This interpretation may be useful, e.g., for modeling of silver dendrites that spread in the [112] direction  \citep{radmilovic2016}, which are frequently termed feather-like structures in the literature.

\subsection{Simulations in HCP lattices}

\begin{figure}[!ht]
	\centering
	\includegraphics[width=0.45\textwidth]{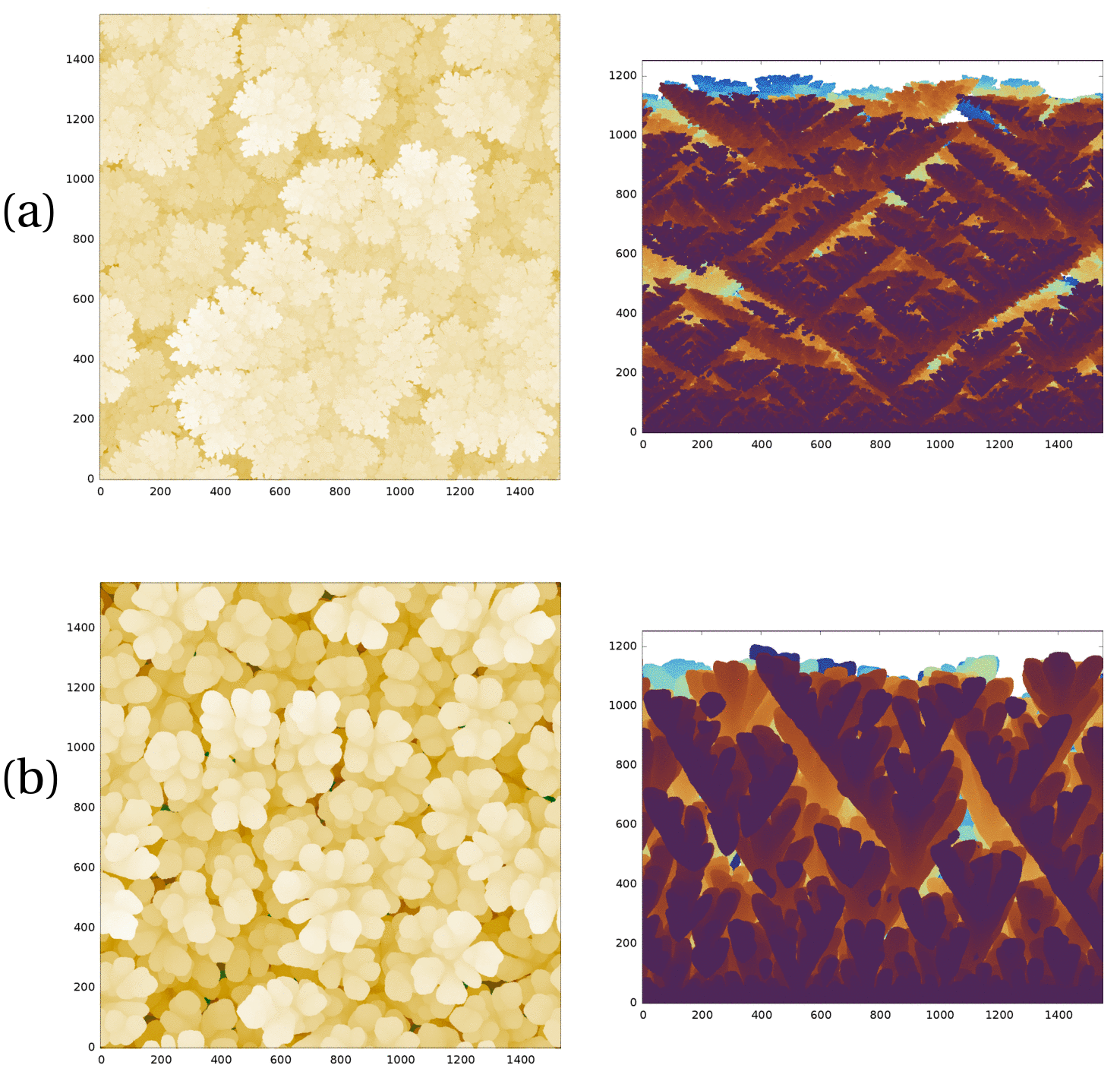}
	\caption{Top and cross-sectional views of deposits grown with: (a) $G\sim{10}^4$, $P=0.2$; (b) $G\sim{10}^4$, $P=0.4$. Adapted from Ref. \citenum{capriotalebreis2025} (CC-BY 4.0 licence).
		}
	\label{examplesHCP}
\end{figure}

Simulations of the model on HCP lattices with nucleation at all points of (0001) electrodes lead to formation of leaf-like dendrites or flower-like dendrites depending on the values of $G$ and $P$ \citep{capriotalebreis2025}.
These morphologies are illustrated in the top and cross-sectional view of Figs. \ref{examplesHCP}(a) and \ref{examplesHCP}(b).

For $G\sim{10}^4$ and $P=0.2$ [Fig. \ref{examplesHCP}(a)], the dendrite trunks propagate in a direction that forms an angle with the electrode and whose projection in the electrode has [10$\overline{1}$0] or the other five equivalent directions.
Close inspection of the dendrites show that their top surfaces are in the lowest energy configuration (0001); the other surfaces are (10$\overline{1}$0) or equivalent, which are also low energy configurations where an adatom has $8$ NNs.
As $P$ increases to $0.4$, more rounded dendrites appear, with flower-like shapes whose petals follow the sixfold symmetry [Fig. \ref{examplesHCP}(b)].

\subsection{Scaling of dendrite size}

% R1Q5
The characteristic widths $W$ of the compact and dendritic structures generated with the electrodeposition model were calculated from cross-sections of the deposits parallel to the electrodes at $\approx2/3$ of the height of the simulation cell \citep{capriotalebreis2018,dokhan2022,carrascoPCCP2022,capriotalebreis2025}.
In the SC lattice, the areas $A$ of the NN-connected domains in those cross sections were measured and $W$ was estimated as $\langle A\rangle^{1/2}$ \citep{capriotalebreis2018,carrascoPCCP2022}.
The results obtained in cross sections at other heights (near $1/2$ of the maximal heights) are similar \citep{capriotalebreis2018}.
In the FCC and HCP, $W$ was defined as the average length of NN-connected segments \citep{dokhan2022,capriotalebreis2025}.

\begin{figure}[!ht]
	\center
	\includegraphics[clip,width=0.32\textwidth,angle=0]{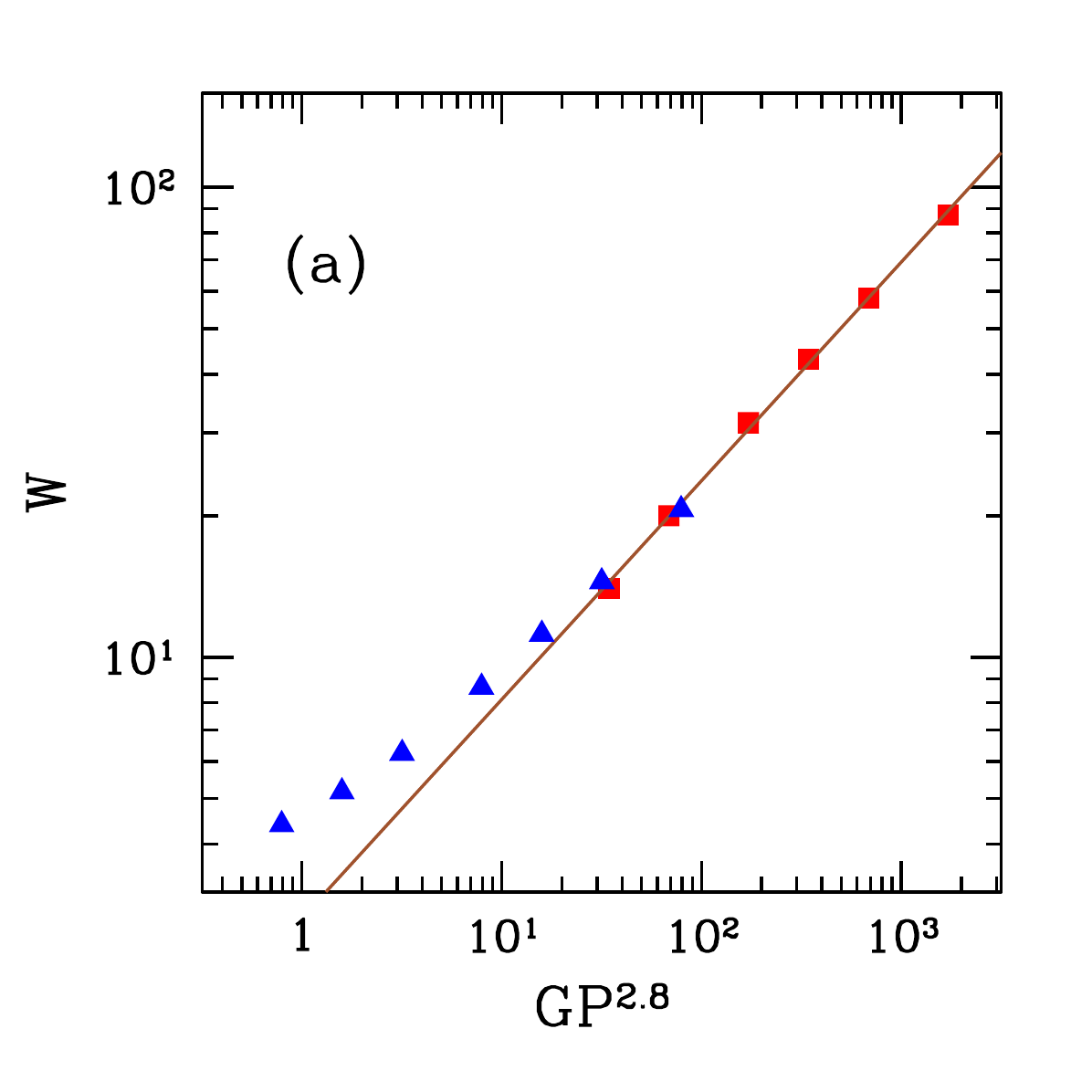}
	\includegraphics[clip,width=0.32\textwidth,angle=0]{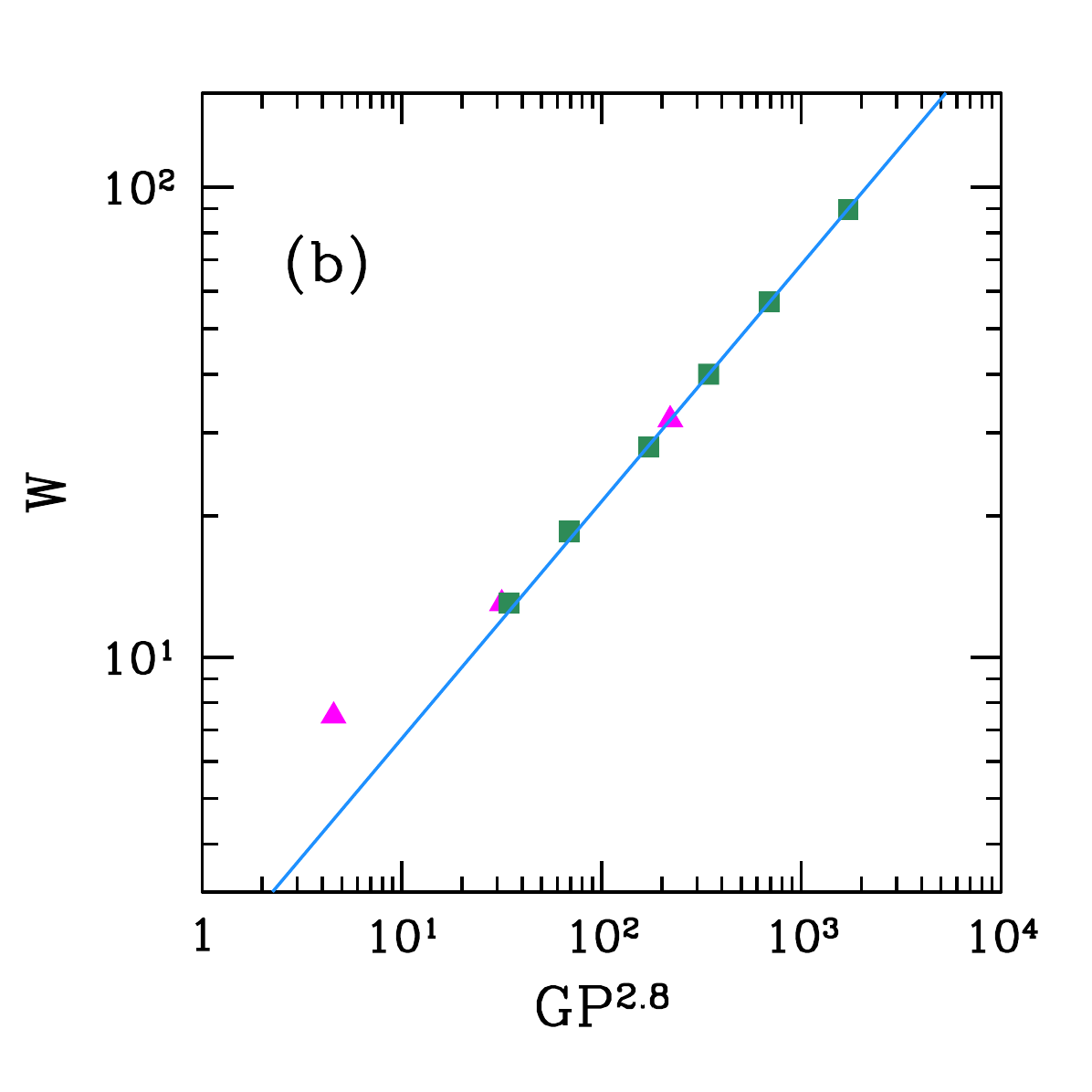}
	\includegraphics[clip,width=0.32\textwidth,angle=0]{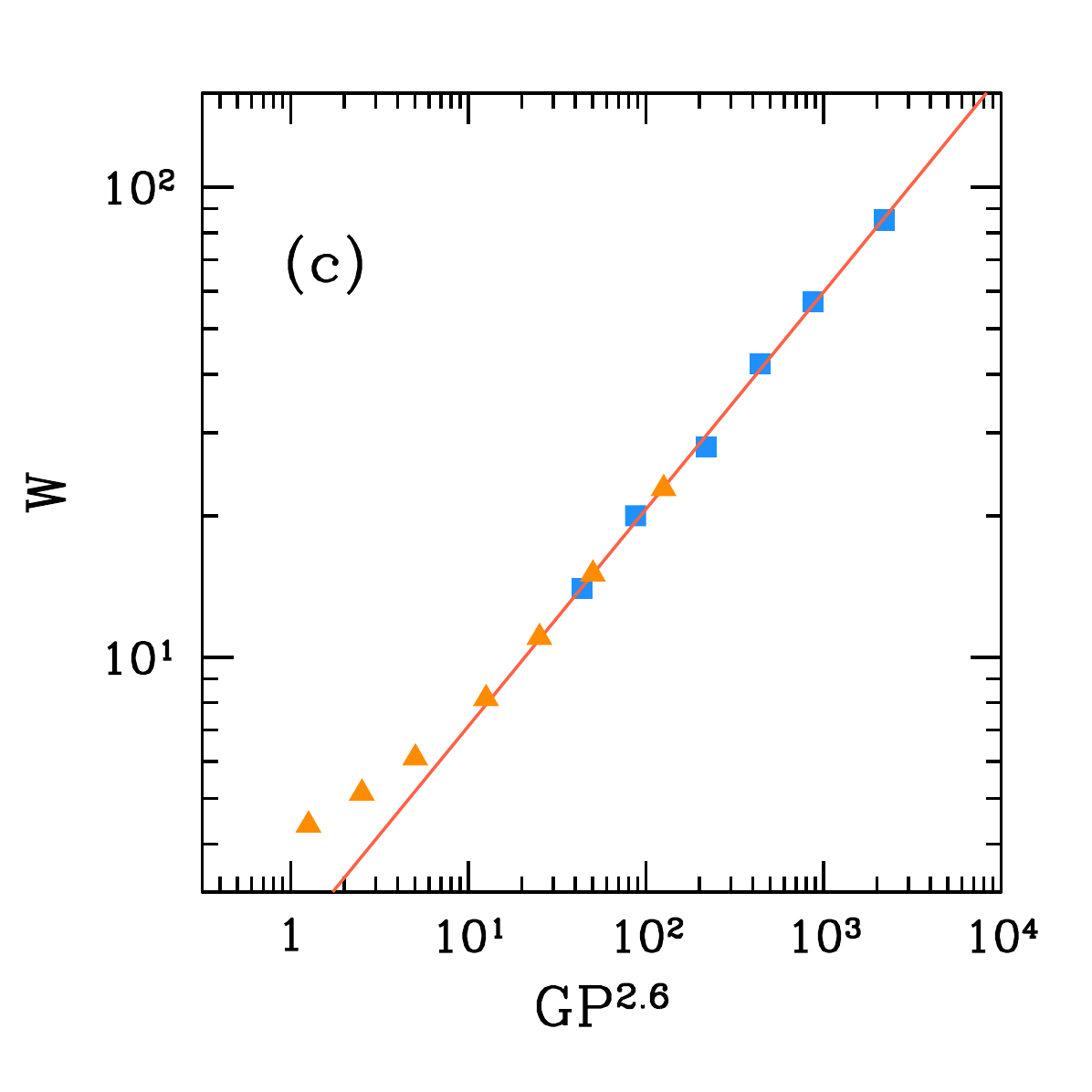}
	\caption{Average dendrite width (in units of the lattice constant) as a function of scaled variables of $G$ and $P$:
		(a) FCC(111) substrates [$P=0.1$ (blue triangles) and $P=0.3$ (red squares)];
		(b) FCC(110) substrates [$P\leq0.2$ (magenta triangles) and $P=0.3$ (green squares)];
		(c) HCP(0001) substrates [$P=0.1$ (orange triangles) and $P=0.3$ (blue squares)].
		Straight lines are linear fits of the data with abscissa $\gtrsim{10}^2$.
		Reproduced from Ref. \citenum{capriotalebreis2025} (CC-BY 4.0 licence).}
	\label{scalingW}
\end{figure}

% R1Q5
Inspection of the deposits in all lattices shows that $W\sim1$ is typically characteristic of DLA-like deposits, while dendrites (i.e., hierarchically organized branched structures) are typically formed when $W\gtrsim10$ (in lattice units).
In this range, quantitative analysis showed that $W$ was generally a function of a combination of the model parameters in the form
\begin{equation}
	W\sim {\left(GP^m\right)}^k ,
	\label{WGP}
\end{equation}
{with positive exponents $m$ and $k$.
For each lattice and orientation, $m$ was selected to collapse data obtained with different values of $P$, whereas $k$ was obtained from the slope of the corresponding double-logarithmic plots.}
The exponent $k$ is generally between $0.4$ and $0.5$ and the exponent $m$ is highly dependent on the lattice structure: $m\approx3$ in FCC and HCP lattices; $m\approx1$ in SC lattices \citep{capriotalebreis2018,dokhan2022,carrascoPCCP2022,capriotalebreis2025}.
Figures \ref{scalingW}(a)-(c) show $W$ as a function of $GP^{2.8}$ for growth on FCC lattices with two orientations and $W$ as a function of $GP^{2.6}$ for growth on HCP lattices; the slopes of these double logarithmic plots range from $0.46$ to $0.50$.

The exponent $k$ close to $1/2$ suggests that the width $W$ is of the same order as the adatom diffusion length, which is expected to be proportional to $G^{1/2}$; a similar interpretation is justified in models of deposition from vapor \citep{mozo2020}.
For understanding the effect of $P$ in different lattices, observe that an adatom on a FCC$\left(111\right)$ or HCP$\left(111\right)$ plane has $3$ NNs belonging to that plane, while an adatom on a SC$\left(001\right)$ plane has $1$ NN belonging to that plane.
Thus, considering the hop probability in Eq. (\ref{Phop}), the factors ${\left(GP^3\right)}^{1/2}$ (for FCC and HCP) and ${\left(GP\right)}^{1/2}$ (for SC) are of the order of the diffusion length of adatoms on the respective lowest-energy planes with $n=1$ lateral NN (i.e. one neighbor on the same plane).
This means that a mobile atom can travel distances of order $W$ on the low-energy terraces even if this adatom is sometimes bound to another adatom on the same terrace, e.g., at the borders of compact islands that are formed on that terrace. 
The adatoms with larger numbers of NNs are effectively frozen, such as those belonging to compact islands or those of the lowest-energy planes (the latter with much larger numbers of NNs) \citep{capriotalebreis2025}.

% R1Q5 - paragraph improved
Observe that the linear fits in Figs. \ref{scalingW}(a)-(c) include only data with $GP^m\gg1$ and $W>10$ (in lattice units).
This means that, in the conditions where dendrites are formed and where the above scaling relations are valid, the number of executed adatom hops is typically large.
The scaling relations fail when the average number of executed hops is of order $1$ or smaller, i.e. $GP^m\lesssim1$, which are the conditions where the random ramification characteristic of DLA is obtained.

% R1Q3
The above relations provide a phenomenological mapping between morphological features and model parameters.
An individual morphological property often constrains the value of $GP^m$, so different $(G,P)$ pairs may produce similar morphologies.
Thus, estimates of these parameters based on a single property are not necessarily unique.
Better-constrained inverse mapping requires measurements of multiple morphological properties; formerly, this was possible only in a study of silver electrodeposition (Sec. \ref{silverapplication}).
Additional difficulties will appear if one tries to estimate the effective parameters of the temperature-activated forms proposed for $G$ and $P$ in Eq. \ref{defGP}.

%R1minor1,3,4
\subsection{Summary of morphology regimes}

Table \ref{tab_morphologies} summarizes the morphology regimes illustrated above for the different lattice structures and presents the representative model parameters.
There is no sharp transition between the morphology regimes, but only smooth crossovers, so the morphological classification is qualitative and it is based on visual inspection of the simulated deposits.
The regimes may be associated with properties such as branch orientation, hierarchy, and rotational symmetry.

\begin{table*}[t]
	\centering
	\caption{Summary of the model's main morphology regimes.}
	\label{tab_morphologies}
	\footnotesize
	\setlength{\tabcolsep}{3pt}
	\renewcommand{\arraystretch}{1.12}
	\begin{tabular}{@{}
			p{0.07\textwidth}
			p{0.16\textwidth}
			p{0.13\textwidth}
			p{0.5\textwidth}
			@{}}
		\hline
		Lattice (electrode orientation) &
		Representative Regime &
		Morphology &
		Structural Feature\\
		\hline
		
		All&
		$GP^m\lesssim1$ ($m\approx1$ in SC, $m\approx3$ in FCC and HCP)&
		DLA-like/fractal &
		Random ramification without repeated branch hierarchy [Fig.~\ref{examplesSC}(a)] \\
		
		SC(001)&
		$G\gtrsim10^4$ and $P\gtrsim0.1$ (Ref. \citenum{capriotalebreis2018}) &
		Compact to rounded-columnar &
		Dense initial layer followed by rounded columns [Fig.~\ref{examplesSC}(b)] \\
		
		SC(001)&
		Intermediate $G$ and $P$ (Ref. \citenum{capriotalebreis2018}) &
		Maple-leaf dendrites &
		Hierarchical branching with tips propagating along the $\langle111\rangle$ directions [Figs.~\ref{examplesSC}(c)-(e)] \\
		
		FCC(001)&
		$GP^3\gtrsim1$, $GP^6\lesssim1$ (Ref. \citenum{capriotalebreis2025}) &
		Pine-tree dendrites &
		Trunks along $[001]$ and secondary branches along the in-plane
		$\langle100\rangle$ directions [Fig.~\ref{examplesFCC}(a)] \\
		
		FCC(111)&
		$GP^3\gtrsim1$, $GP^6\lesssim1$ (Ref. \citenum{capriotalebreis2025}) &
		Leaf-like dendrites &
		Trunks along $[001]$, $[010]$, and $[100]$, with fewer secondary
		branches [Fig.~\ref{examplesFCC}(b)] \\
		
		HCP(0001)&
		$GP^3\gtrsim1$, $GP^6\lesssim1$ (Ref. \citenum{capriotalebreis2025}) &
		Leaf-like dendrites &
		Inclined trunks whose projections follow $[10\bar{1}0]$ and the other equivalent directions [Fig.~\ref{examplesHCP}(a)] \\
		
		HCP(0001)&
		$GP^6\gtrsim1$ (Ref. \citenum{capriotalebreis2025}) &
		Flower-like dendrites &
		Rounded petals displaying sixfold symmetry [Fig.~\ref{examplesHCP}(b)] \\
		\hline
	\end{tabular}
\end{table*}

\section{Applications}
\label{applications}

\subsection{Silver electrodeposition on a pattern of gold nanoparticles}
\label{silverapplication}

Dendritic silver has applications in catalysis and SERS \citep{cheng2019,raj2020}, in superhydrophobic coatings \citep{xuLang2010}, and as antibacterial materials \citep{palAEM2007}.
The first direct application of the model was silver electrodeposition on (0001) highly oriented pyrolitic graphite (HOPG) electrodes modified with $4$~nm Au nanoparticles (NPs) \citep{dokhan2022}.
The uniform coverage provided by these NPs allowed nucleation throughout the entire electrode area and, consequently, the growth with nearly constant current since early times.

\begin{figure}[!h]
	\centering
	\includegraphics[width=0.5\textwidth]{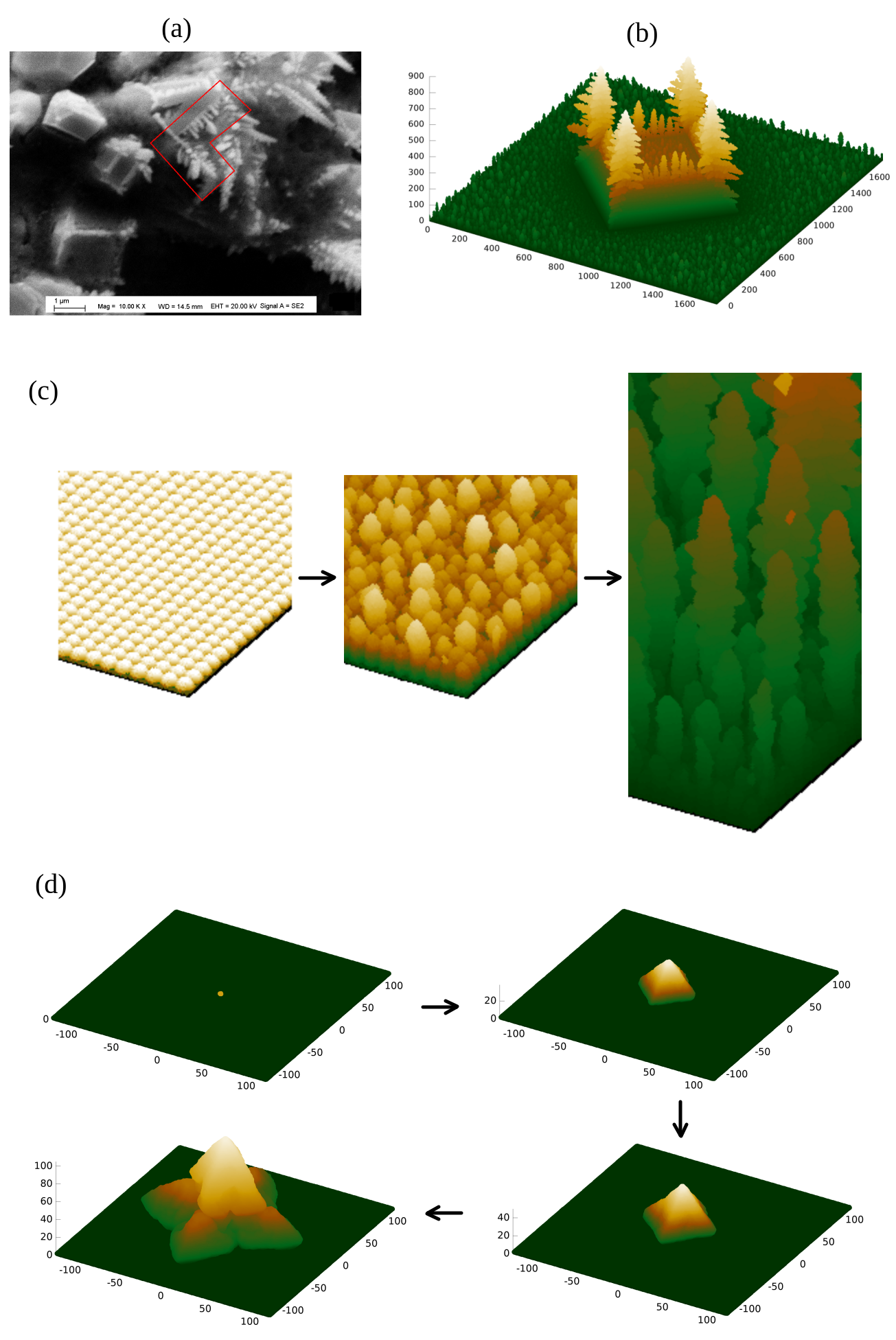}
	\caption{(a) Highlighted nanotrees growing from vertices and edges of silver microparticles in electrodeposition on Au NP HOPG substrates.
		(b) Simulated deposit on prebuilt faceted particles with $G={10}^4$ and $P=0.1$.
		(c) Evolution of the deposit on a patterned NP surface obtained in simulations with $G={10}^4$ and $P=0.1$.
		(d) Evolution of the deposit on a single NP seed obtained in simulations with $G={10}^6$ and $P=0.05$. Adapted with permission from Ref. \citenum{dokhan2022}. Copyright 2022 American Chemical Society.}
	\label{figsilver}
\end{figure}

Figure \ref{figsilver}(a) shows a SEM image of a final deposit obtained after $\approx600$~s of operation.
Faceted silver microparticles with dominant $\left(111\right)$ orientation are observed; they are much larger than the Au NPs, which are not visible at this scale.
Moreover, dendrites with a pine tree shape grow from the vertices and edges of those microparticles.
A small decrease of the current (less than $20\%$) is observed between $\approx300$~s and $\approx600$~s of operation; this feature is associated with the dendrite nucleation and growth.
These results represent a morphological transition from an initial compact morphology to a final dendritic morphology.

%R1Q4
Simulations of the model on a FCC lattice considering large prebuilt pyramidal particles on a flat electrode reproduce the preferential growth of dendrites at vertices and edges, as shown in Fig. \ref{figsilver}(b).
This feature is interpreted as a tip effect because the most protuberant parts of the deposit are more easily reached by the cations that are released well above the deposit and move in a highly depleted region of the electrolyte.
As the dendrites grow, this tip effect is enhanced, and the cations seldom reach the surface of the original particle or the bottom parts of the dendrites.

Figure \ref{figsilver}(c) shows simulations with the evolution of a deposit considering a $\left(001\right)$ electrode patterned with hemispherical NPs (left panel), which mimics the electrodeposition HOPG electrodes modified with $4$~nm Au NPs.
Small pyramidal particles are initially formed (middle panel), while dendrites with pine tree shape appear at longer times (right panel).
Fig. \ref{figsilver}(d) shows simulation results with a single small seed that illustrate the formation of larger pyramids with $\left(111\right)$ surfaces [despite the $\left(001\right)$ orientation of the underlying plane] followed by branching that will give rise to dendrites.
Thus, the morphological transition observed in the experiments is qualitatively reproduced by the model.

However, these simulations are performed in the accessible sizes of simulation cells, which are roughly of $400$~nm, so the compact particles and dendrites have nanoscale sizes, which are well below the characteristic sizes of the electrodeposited silver.
A quantitative description of the experimental data, involving much larger sizes, required two steps: (i) the estimation of the sizes of the compact and of the dendritic structures for several values of the model parameters ($G$ and $P$); (ii) the extrapolation of these results to the realistic size range.

In step (i), the average dendrite width $W$ measured in simulations with $(001)$ substrate was shown to follow the relation $W \approx 3.6{\left(GP^3\right)}^{0.39}$, with $W$ in lattice units.
The sizes $S_t$ of the compact particles in the transitions to the dendritic morphology were determined from pyramid sizes such as that shown in Fig. \ref{figsilver}(d), specifically considering the point where there were significant deviations in the ratio between the diagonal and the lateral lengths of pyramid bases (expected ratio is $\sqrt{2}$).
For $0.01\leq P\leq0.05$, the approximate dependence on the model parameters is $\log_{10}{S_t} \approx \left( 0.51-4.7~P\right) \log_{10}{G} + 39~P-1.8$, with $S_t$ in lattice units.
Interestingly, for $P<0.27$ and sufficiently large $G$, $S_t$ increases faster with $G$ than $W$ does, indicating that compact particles tend to be larger than the dendrites.
This is actually observed in the experiments.

The electrodeposition of silver shows microparticles with facets of maximal sizes $S_{exp}\sim1\mu\text{m}$ and nanotrees with characteristic sizes $W_{exp}\approx350$nm; the latter were averages between trunk widths and secondary branch lengths, for consistency with the method used to measure $W$ in the simulations.
The above relations for the model fit these values with $G=4\times{10}^{11}$ and $P=0.03$; as expected, a value of $G$ much larger than those considered in the simulations is obtained.

From these parameter values, the estimated diffusion lengths of free adatoms on $(111)$ and $(001)$ terraces were of the same orders as $S_{exp}$ and of the average trunk width in the experiments.
The $(001)$ terraces are formed at the tip of the dendrites as they propagate in this (or equivalent) directions.
Thus, this combination of modeling and experimental data is consistent with adsorbate diffusion determining the characteristic widths of the electrodeposited material at short length scales. %Referee 1 major 4

%R2Q1 , R2Q2
% R1Q2 complement
Another interesting consequence of these results is that the diffusion lengths of Ag atoms (from their reactions to their permanent aggregation) are nearly the same on $(001)$ and $(111)$ surfaces, in striking contrast with Ag diffusion on the same surfaces during vapor deposition of this material \citep{etb,Nelli2023}. 
This comparison for a specific material helps to highlight the differences between these techniques and between their modeling approaches.
During vapor deposition, the motion of the adatoms on a film surface determines the sizes and shapes of islands or mounds in each crystalline orientation, while the corresponding models approximate that motion with orientation-dependent diffusion coefficients to reproduce the measured sizes and shapes.
Instead, electrodeposition occurs at an electrified solid-liquid interface where solvation, applied potential, and other mechanisms lead to very different relaxation processes, such as the relaxation of amorphous or polycrystalline layers observed in TEM of silver dendrites \citep{fang2006}.
The estimated diffusion lengths of Ag atoms on $(001)$ and $(111)$ surfaces are consequences of this complex relaxation.
Thus, the minimal model presented here has effective parameters that provide average sizes of compact particles and dendrites consistent with the experiments, but which are not approximations of the individual motion of the adatoms.

\begin{figure*}%[t]
	\centering
	\includegraphics[width=\textwidth]{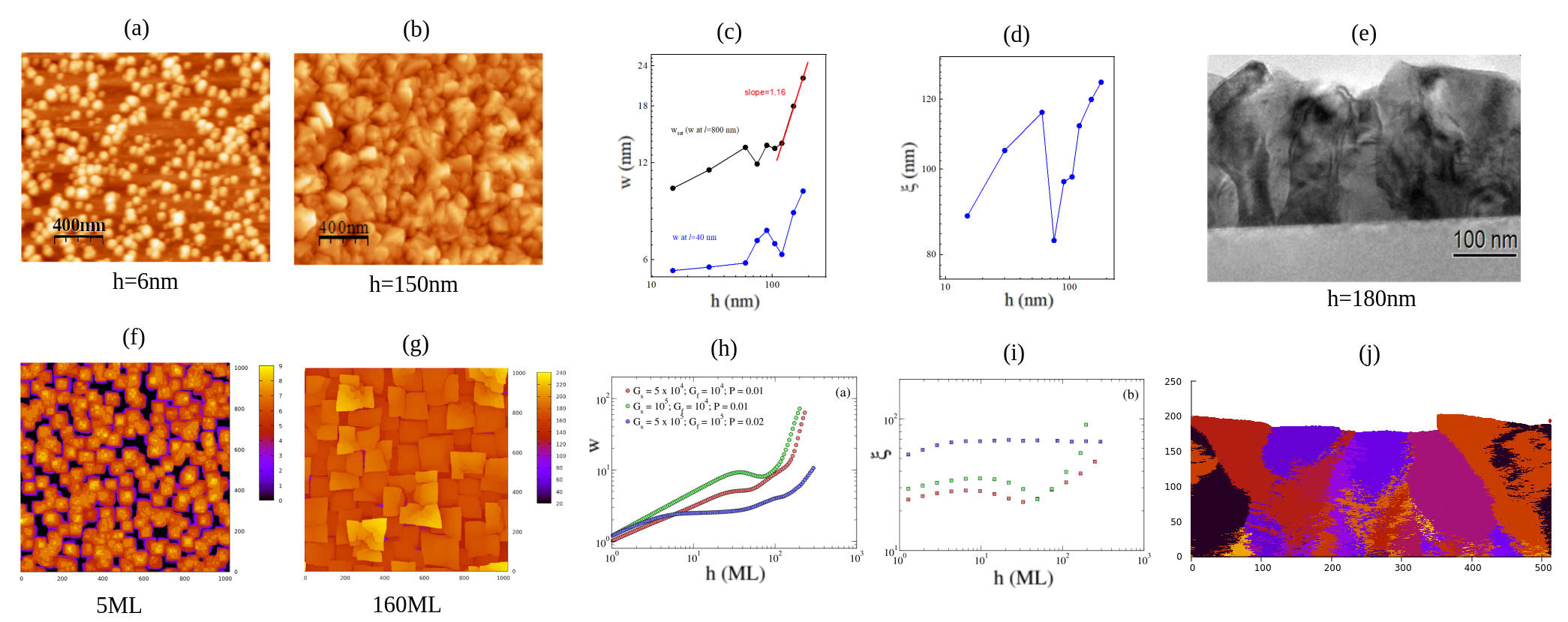}
	\caption{(a),(b) AFM images of the indicated thicknesses of electrodeposited Fe on {\it n}-Si(100).
		(c) Roughness as a function of the thickness measured in entire AFM images ($l=800$~nm) and averaged over gliding boxes of sizes $l=40$~nm.
		(d) Lateral correlation length as a function of the thickness obtained from AFM images.
		(e) TEM image of an electrodeposited Fe film with a thickness of $180$~nm.
		(f),(g) Top views of deposits obtained in simulations with $G_s={10}^5$, $G_f={10}^4$, and $P=0.01$, with the indicated thicknesses in monolayer (ML).
		(h) Roughness as a function of the thickness obtained in simulations with three parameter sets.
		(i) Lateral correlation length as a function of the thickness obtained in simulations with the same parameter sets.
		(j) Cross section of a deposit obtained in simulations with the same parameters as in (f) and (g) and the indicated thickness.
		Adapted with permission from Ref. \citenum{benetti2023}. Copyright 2023 American Chemical Society.}
	\label{figFe}
\end{figure*}

\subsection{Application to iron electrodeposition on Si}

Iron is ferromagnetic at room temperature, is abundant in nature, and is environmentally friendly, so the integration of thin iron films with semiconductors may be a step towards the substitution of heavy and rare elements for magnetic applications \citep{chenACSAMI2015}.
This motivated a study of Fe electrodeposition on {\it n}-Si(100) wafers and the extension of the above modeling approach that accounts for different nucleation mechanisms \citep{benetti2023}.
The native oxide was removed from the Si substrate and the electrodeposition was carried out potentiostatically at $1.50$~V relatively to a saturated calomel electrode, for nearly $9$~min, and with an electrolyte solution containing $\text{FeSO}_4$, $\text{H}_3\text{BO}_3$, and $\text{H}_2\text{SO}_4$ to adjust the pH to $2.3$.

Fig. \ref{figFe}(a) shows an atomic force microscopy (AFM) image of an initial stage of the electrodeposition, in which
non-uniform nucleation leads to the spread of multilayer islands.
The applied current increased from $3~\text{mA/cm}^2$ at the beginning of the process to $4.5~\text{mA/cm}^2$ when this $6$~nm sample was formed, but rapidly dropped and was maintained in the range $3\text{--}3.5~\text{mA/cm}^2$ for thicknesses varying from $\sim15$~nm to $\sim150$~nm.
The thickest films have a grainy surface morphology, as shown in Fig. \ref{figFe}(b).
The surface roughnesses $w$ shown in Fig. \ref{figFe}(c) were measured in the entire AFM images (size $l=800$~nm) and averaged over gliding boxes of sizes $l=40$~nm.
Both measures show oscillations at intermediate thicknesses.
The lateral correlation length $\xi$, shown in Fig. \ref{figFe}(d), has a deeper minimum correlated with the roughness oscillation.

A TEM image of the thickest ($180$~nm) sample shown in Fig. \ref{figFe}(e) reveals the growth of a polycrystalline film.
TEM dark-field images of this sample and of intermediate growth stages showed individual single crystalline grains and revealed secondary nucleation, i.e., the formation of new grains on the surface of the previously deposited Fe grains.
The bottom of the secondary grains were located $50\text{--}150$~nm above the Si substrate, so their appearance correlates with the decrease of the lateral correlation length and with the roughness fluctuations.
Since these grains begin to grow with small widths, the consequent surface undulations have small lateral extent and explain the decrease in $\xi$.

%R1Q7
An extended version of the electrodeposition model reproduces the initial island formation and the roughening of these Fe films \citep{benetti2023}.
The cations were assumed to react immediately in contact with the deposited Fe, but this reaction had a probability ${10}^3$ times smaller in contact with the Si electrode.
Moreover, the diffusion of Fe atoms on the Si substrate was assumed to be faster than that on the Fe film, which is represented by different numbers of allowed hops $G_s$ and $G_f$ in the respective regions, with $G_s>G_f$.
This is expected due to the weaker interaction between the different materials (Fe and Si).
Recent studies of inorganic and organic films deposited from vapor showed that these conditions lead to the formation of multilayer islands independently of the lattice structure \citep{toApplSS2021,lapkin2026}.
However, the polycrystallinity of the Fe films grown on Si(100) cannot be represented in lattice models where a single crystalline orientation is defined for the whole deposit.
This limitation implies that only qualitative descriptions of the Fe film evolution are expected, so a SC lattice was adopted to simplify the simulation work.

Figures \ref{figFe}(f) and \ref{figFe}(g) show images of deposits obtained in the simulations of this model with $G_s={10}^5$, $G_f={10}^4$, and $P=0.01$.
The $5$~ML deposit shows the initial formation of multilayer islands, in parallel with the AFM result in Fig. \ref{figFe}(a), and the rough pattern in the $160$~ML film shows a grainy structure that resembles the AFM image of Fig. \ref{figFe}(b).
However, observe that the scales of the simulated and of the experimental images differ because iron is a BCC metal and its lattice constant is $\sim0.29$~nm.

The evolution of the roughness $w$ (in ML units) obtained in simulations with several parameter sets is shown in Fig. \ref{figFe}(h).
The first maximum of the roughness (or the onset of a plateau for $G_f={10}^5$) indicates the beginning of the coalescence of the initial islands, which leads to the formation of a continuous film with possibly smaller roughness.
However, in the simulations, the diffusive cation flux favors reduction and adatom aggregation at the most protuberant parts of the surface, leading to the subsequent instability and rapid increase in roughness. %Referee 1 major 4
In the two data sets with $G_f={10}^4$ in Fig. \ref{figFe}(h), the small oscillations of the roughness followed by its sharp increase are similar to the experimental result in Fig. \ref{figFe}(c); the above-mentioned difference in the scales of the simulated and of the experimental data is maintained.

The correlation length $\xi$ obtained in simulations with the same parameter sets is shown in Fig. \ref{figFe}(i).
The data for $G_f={10}^4$ show small oscillations in the thickness region where island coalescence occurs.
These fluctuations are quantitatively smaller than those in the experiments [Fig. \ref{figFe}(d)], but this is expected because the model did not include the renucleation mechanism.

Figure \ref{figFe}(j) shows a cross-section of a simulated deposit in which the different colors distinguish the grains formed from the initial islands (using the algorithm of Ref. \citenum{reisPRE2017}).
The surface landscape resembles the experimental one [Fig. \ref{figFe}(e)], with the particular feature that the high surface steps which lead to instability development are located near the grain boundaries. 

\section{Conclusions}
\label{conclusion}

The results of simulations of an electrodeposition model incorporating diffusive cation flux, very fast reaction, and adsorbed atom diffusion in finite time windows were reviewed.
In flat electrodes and when the diffusion lengths of the adsorbed atoms are of $\sim10$ lattice units or more, compact films are initially formed, but instabilities eventually develop and give rise to dendrite growth, i.e., hierarchical structures in which branches of a given order reproduce the shape and crystalline orientation of the branches of the previous order.
Instead, for diffusion lengths of the same order or smaller than the lattice constant, fractal deposits similar to DLAs are formed.
When the nucleation is restricted to a nanosized seed, a compact particle is formed and a crossover to dendrite growth is observed, which constitutes a dynamic trasition similar to that observed in silver electrodeposition on substrates patterned with gold nanoparticles.

The dendrite shapes depend on the lattice structure and on the orientation of the cation flux because the latter is able to suppress growth of some secondary and tertiary branches characteristic of that lattice.
The dendrite surfaces are predominantly oriented in the lowest-energy configurations of each lattice, and the average dendrite width is proportional to the diffusion lengths of adsorbed atoms weakly bound to islands growing on these lowest-energy planes.
These morphologies result from the competition between two processes: the relaxation that produces low-energy configurations at short length scales and the diffusive cation flux that privileges reaction at protuberant surface points and leads to unstable growth at long length scales.

A quantitative application to the transition between faceted silver microparticles and dendritic silver leads to values of the model parameters which consistently provide diffusion lengths of atoms on (111) and (001) surfaces of the same order as the characteristic widths of those structures.
The simulations of an extended model for iron electrodeposition on Si(100) reproduce the initial formation of multilayer islands, small oscillations of the roughness of a polycrystalline compact film after island coalescence, and instability development after $100$~nm of thickness. %Referee1 major 4

These studies suggest possible directions for applying the model to the electrodeposition of other materials.
In silver and copper electrodeposition, some microscopy studies show the formation of dendrites with shapes of feathers or leaves that differ from those shown in previous simulations \citep{talebEActa2013,talebEActa2013a,radmilovic2016}.
The difference may be a consequence of the simple features of the model presented here, which has a simple energetic interpretation: since $P$ may be expressed as a Boltzmann factor [Eq. (\ref{defGP})], the activation energy for a hop is proportional to the number $n$ of occupied NNs [Eq. (\ref{Phop})], i.e., the activation energy is additive over the NNs.
Instead, there may be changes in the dendrite shapes if one accounts for an orientation dependence of the atom-atom interactions, which affects the diffusion on different planes and does not obey the additive rule.
Such a dependence is recognized to be important for modeling surface diffusion during vapor deposition \citep{etb}.
Another possible direction for future studies is modeling the electrodeposition of lithium dendrites.
With the present model or with some of its extensions, simulations in body centered cubic lattices may be the first step.
Finally, current studies of the model aim at determining the relation between the growth conditions and the accessible surface area of dendrites, which may be relevant for SERS applications \citep{cervantes2026}.

\section*{Conflicts of interest}
There are no conflicts of interest to declare.

% R1Q8
\section*{Data availability}

All data presented here were published in Refs. \citenum{reiscapriotaleb2017}, \citenum{capriotalebreis2018}, \citenum{dokhan2022}, \citenum{benetti2023}, and \citenum{capriotalebreis2025}.
These data are available by request to the authors.

\section*{Acknowledgements}

FDAAR acknowledges support from the Brazilian agencies FAPERJ (E-26/202941/2015, E-26/202.881/2018, E-26/210.354/2018,E-26/
210.062/2023, E-26/201.050/2022), CNPq (304766/2014-3, 305391/2018-6, 442233/2023-0, 305570/2022-6), and CAPES (88881.068506/2014-01, 88887.310427/2018-00, 88881.700849/2022-01).
SGA acknowledges support from the Brazilian agencies CNPq (405864/2021-3, 	306094/2025-8) and FAPEMIG (APQ-00382-16).
ISSC acknowledges support from the Brazilian agency FAPERJ (E-26/202.355/2018), CNPq (408929/2025-1).

\balance

%%%REFERENCES%%%
%\bibliography{interfaces} %You need to replace "rsc" on this line with the name of your .bib file
%\bibliographystyle{rsc} %the RSC's .bst file

\providecommand*{\mcitethebibliography}{\thebibliography}
\csname @ifundefined\endcsname{endmcitethebibliography}
{\let\endmcitethebibliography\endthebibliography}{}

\end{document}